\documentclass[11pt]{article}

\usepackage{hyperref,amsfonts,amsmath,amssymb,graphicx,bm,subfigure,a4wide,cite}

\numberwithin{equation}{section}

\begin{document}

\title{\textbf{Low mode approximation in the axion magnetohydrodynamics}}

\author{Maxim Dvornikov\thanks{maxim.dvornikov@gmail.com}
\\
\small{\ Pushkov Institute of Terrestrial Magnetism, Ionosphere} \\
\small{and Radiowave Propagation (IZMIRAN),} \\
\small{108840 Moscow, Troitsk, Russia}}

\date{}

\maketitle

\begin{abstract}
We study the evolution of interacting large scale magnetic and axionic
fields. Based on the new induction equation accounting for the contribution
of spatially inhomogeneous axions, we consider the evolution of a
magnetized spherical axion structure. Using the thin layer approximation,
we derive the system of the nonlinear ordinary differential equations
for harmonics of poloidal and toroidal magnetic fields, as well as
for the axion field. In this system, we account for up to four modes.
Considering this small and dense axion clump to be in a solar plasma,
we numerically simulate the evolution of magnetic fields. We obtain
that the behavior of magnetic fields depends on the initial fields
configuration. Moreover, we find an indication on a magnetic field
instability in the magnetohydrodynamics with inhomogeneous axions.
\end{abstract}

\maketitle

\section{Introduction}

Nowadays it is commonly accepted that a significant fraction of the
universe mass is present in the form of dark matter, which reveals
itself mainly by the gravitational interaction. Despite multiple candidates
for dark matter particles were proposed, the nature of dark matter
is still unclear. The most plausible candidates for dark matter constituents
are light particles called axions~\cite{Luz20}, which were first
proposed in the quantum chromodynamics (QCD)~\cite{PecQui77,Wei78,Wil78}. The
masses of QCD axions are in the range $10^{-6}\,\text{eV}<m<10^{-4}\,\text{eV}$.
However, much lighter axion like particles (ALPs) with masses $m<10^{-11}\,\text{eV}$
are also considered~\cite{Mar16,Cho21}. In spite of numerous attempts
aimed to the direct detection of axions and ALPs in a laboratory~\cite{IraRed18,Sik21},
these particles are still elusive.

Despite dark matter interacts mainly gravitationally, axions and ALPs
are admitted to interact with electromagnetic fields, although the
corresponding coupling constant is quite small. The interaction between
dark and usual matter results in various interesting phenomena like
the emission of ultra-high-energy cosmic rays~\cite{Iwa00} and various
multi-messenger effects~\cite{Die19}. Moreover, gamma ray bursts
with certain energies are explained in Ref.~\cite{Mul23} by decays
of ALPs.

Axions background proceeding from the early universe is likely to
be spatially homogeneous. However, if the Peccei--Quinn (PQ) phase
transition happens after the reheating during the inflation, inhomogeneities
of axions can be present. They can result in the formation of axion
miniclusters~\cite{KolTka93,KolTka94} and axion stars~\cite{ChaDel11,BraZha19}.

The electrodynamics equations in presence of axions were obtained in Refs.~\cite{Sik83,Wil87} to describe possible experiments to detect these particles.
The interaction of magnetic fields with spatially homogeneous axions
were shown in Ref.~\cite{LonVac15} to cause the magnetic field instability.
We considered the magnetohydrodynamics (MHD) in the presence of spatially
homogeneous~\cite{DvoSem20} and inhomogeneous~\cite{Dvo22} axions
in the early universe. Besides the magnetic field instability the
parametric instability is reported in Ref.~\cite{Bey23} to take
place in the axion electrodynamics.

Recently, in Refs.~\cite{DvoAkh24,AkhDvo24}, we derived the new
induction equation for the magnetic field evolution in presence of
inhomogeneous axions and applied it for magnetic fields with various
configurations including quite nontrivial ones related to the Hopf
fibration. This new equation was used in Ref.~\cite{Dvo24} to study
the magnetic field behavior in a dense axion star in solar plasma.
A possible implication of the obtained results to the solar corona
heating was discussed in Ref.~\cite{Dvo24}. The axion dynamo in
a neutron star was considered in Ref.~\cite{Anz23}. In the present
paper, we develop the research initiated in Ref.~\cite{Dvo24}.

This work is organized in the following way. First, in Sec.~\ref{sec:AXMHD},
starting with the new induction equation accounting for the axions
inhomogeneity, we derive the main equations describing mutual evolution
of axion and magnetic fields. Then, we formulate the initial condition
and fix the rest of the parameters of the system. In Sec.~\ref{sec:RES},
we present the results of numerical simulations of the magnetic fields
evolution inside a spherical axionic clump in solar plasma. Finally,
we conclude in Sec.~\ref{sec:CONCL}. The ordinary differential equations
for the harmonics of axion and magnetic fields are derived in Appendix~\ref{sec:ODEHARM}.

\section{MHD in presence of axions\label{sec:AXMHD}}

The new induction equation, accounting for the time dependent and
spatially inhomogeneous axion field $\varphi(t,\mathbf{x})$, has
the form~\cite{DvoAkh24,AkhDvo24},
\begin{equation}\label{eq:indeq}
  \frac{\partial\mathbf{B}}{\partial t}=\nabla\times
  \left[
    \mathbf{b}\times(\nabla\times\mathbf{B})+\alpha\mathbf{B}
    -\eta(\nabla\times\mathbf{B})
  \right],
\end{equation}
where $\alpha=g_{a\gamma}\eta\partial_{t}\varphi$ is the $\alpha$-dynamo
parameter, which is nonzero for the time dependent $\varphi$, $\mathbf{b}=g_{a\gamma}\eta^{2}\nabla\varphi$
is the axial vector being nonzero in case of the spatial inhomogeneous
$\varphi$, $\eta$ is the magnetic diffusion coefficient, and $g_{a\gamma}$
is the axion-to-photon coupling constant. We should add to Eq.~(\ref{eq:indeq})
the equation for the axions evolution in presence of a magnetic field.
It is the inhomogeneous Klein-Gordon equation,
\begin{equation}\label{eq:KGeq}
  \frac{\partial^{2}\varphi}{\partial t^{2}}-\Delta\varphi+m^{2}\varphi=
  \eta g_{a\gamma}\mathbf{B}\cdot(\nabla\times\mathbf{B}),
\end{equation}
where $m$ is the axion mass. Note that, in Eqs.~(\ref{eq:indeq})
and~(\ref{eq:KGeq}), we keep only the terms linear in $g_{a\gamma}$.

We use the spherical coordinates $(r,\vartheta,\phi)$ to describe
the evolution of $\mathbf{B}$ and $\varphi$ in Eqs.~(\ref{eq:indeq})
and~(\ref{eq:KGeq}). The magnetic field is taken to have the poloidal
$\mathbf{B}_{p}=\nabla\times(A\mathbf{e}_{\phi})$ and toroidal $\mathbf{B}_{t}=B\mathbf{e}_{\phi}$
components, $\mathbf{B}=\mathbf{B}_{p}+\mathbf{B}_{t}$, where $A$
and $B$ are the new functions of time and coordinates and $\mathbf{e}_{\phi}$
is the unit vector along the $\phi$ variation. We suppose that $A$,
$B$, and $\varphi$ are axially symmetric and have the following
properties in reflection with respect to the equatorial plane:
\begin{equation}\label{eq:symprop}
  A(r,\pi-\vartheta,t)=A(r,\vartheta,t),
  \quad
  B(r,\pi-\vartheta,t)=-B(r,\vartheta,t),
  \quad
  \varphi(r,\pi-\vartheta,t)=-\varphi(r,\vartheta,t),
\end{equation}
 since $A\mathbf{e}_{\phi}$ is the vector, $\mathbf{B}_{t}$ is the
axial vector, and $\varphi$ is the pseudoscalar.

Following Ref.~\cite{Par55}, we assume that the functions $A$,
$B$, and $\varphi$ evolve in a thin layer $R<r<R+\mathrm{d}r$,
where $R$ is the typical radius of the axion structure and $\mathrm{d}r\ll R$
is the layer width. In this approximation, we can neglect the radial
dependence in Eqs.~(\ref{eq:indeq}) and~(\ref{eq:KGeq}), where
we set $r=R$ and $\partial_{r}=0$. Note that we keep the dependence
on the latitude $\vartheta$. The next reasonable approximation is
based on the decomposition over the angular harmonics and keeping
only a few of them. It is called the low mode approximation~\cite{NefSok10}.

Using the dimensionless variables
\begin{equation}\label{eq:dmnlsvar}
  \mathcal{A}=g_{a\gamma}A,\quad\mathcal{B}=g_{a\gamma}RB,
  \quad
  \Phi=\frac{g_{a\gamma}\eta}{R}\varphi,
  \quad
  \tau=\frac{\eta t}{R^{2}},
\end{equation}
we take that
\begin{equation}\label{eq:ABPhiharm}
  \mathcal{A}(\vartheta,\tau) =
  \sum_{k=1}^{4}a_{k}(\tau)\sin(2k+1)\vartheta,
  \quad
  \mathcal{B}(\vartheta,\tau) =
  \sum_{k=1}^{4}b_{k}(\tau)\sin2k\vartheta,
  \quad
  \Phi(\vartheta,\tau) =\sum_{k=1}^{4}\phi_{k}(\tau)\sin2k\vartheta.
\end{equation}
The decomposition in Eq.~(\ref{eq:ABPhiharm}) accounts for the symmetry
properties in Eq.~(\ref{eq:symprop}). Unlike Ref.~\cite{Dvo24},
where only two harmonics were taken into account, we deal with four
harmonics in Eq.~(\ref{eq:ABPhiharm}). As found in Ref.~\cite{Obr23},
in treating solar magnetic fields in frames of the $\alpha\Omega$-dynamo,
one has to account for up to five harmonics.

Based on the decomposition in Eq.~(\ref{eq:ABPhiharm}) and using
the technique in Ref.~\cite{Dvo24}, we obtain the system of nonlinear
ordinary differential equations for the coefficients $a_{k}$, $b_{k}$,
and $\phi_{k}$. Since this system is quite cumbersome, we present it 
in Eq.~(\ref{eq:abpsieq}) in Appendix~\ref{sec:ODEHARM}. In Eq.~(\ref{eq:abpsieq}),
$\mu=mR^{2}/\eta$ and $\kappa=R/\eta$ are is the dimensionless axion
mass and wave vector respectively.

The system in Eq.~(\ref{eq:abpsieq}) is to be solved numerically.
For this purpose we have to specify the initial condition for the
functions $a_{k}$, $b_{k}$, and $\phi_{k}$. To highlight the influence
of the greater number of harmonics on the evolution of magnetic fields
we choose the parameters of the system and the initial condition in
the same form as in Ref.~\cite{Dvo24}. Analogously to Ref.~\cite{Dvo24},
we suppose that an axionic structure exists in the solar plasma.

First, we set the parameters of axions. The axion mass is taken to
be $m=10^{-5}\,\text{eV}$, i.e. we consider a QCD axion. The size
of the axion clump is $R=0.7\,\text{cm}$. The initial energy density
of axions inside such a small axion star is $\rho_{0}\approx10^{-2}m^{2}f_{a}^{2}$,
where $f_{a}\approx5.7\times10^{11}\,\text{GeV}$ is the PQ constant
for the chosen axion mass. Here, we utilize the approximate relation between the axion mass and the PQ constant, $\tfrac{f_a}{10^{12}\,\text{GeV}} \approx 5.7 \left(\tfrac{m}{10^{-6}\,\text{eV}}\right)^{-1}$ (see, e.g., Ref.~\cite{Cha21}). Thus, the axion-to-photon coupling constant, used in Eq.~\eqref{eq:dmnlsvar}, is also uniquely defined by the axion mass, $g_{a\gamma} = \tfrac{\alpha_\mathrm{em}}{2\pi f_a} = 2\times 10^{-15}\,\text{GeV}^{-1}$, where $\alpha_\mathrm{em} = 7.3\times10^{-3}$ is the fine structure constant.

The size and the energy density of the axion clump, used in our simulations, indicate that we deal with a dense axion star. The existence of such an object was proposed in Ref.~\cite{Bra16}, where it was claimed to be stable. However, the analysis in Ref.~\cite{Vic18} showed that dense axion stars decay in a short time interval. The instability of dense stars is because axions in these objects do not have a conserved quantum number. This case is opposite to dilute stars where the quantity $\smallint \mathrm{d}^3 x \varphi^\dagger \varphi$ is conserved (see, e.g., Ref.~\cite{Lev22}).

It was reported in Ref.~\cite{Vic18} that, in some cases, the lifetime of a dense star can reach $\tau_\mathrm{life}\gtrsim 10^{-8}\text{s}$ for $m = 10^{-5}\,\text{eV}$. We shall see shortly in Sec.~\ref{sec:RES} that typical frequencies of magnetic fields oscillations are $f_\mathrm{B} \sim 10^{11}\,\text{Hz}$. Therefore, the period of magnetic oscillations is $\tau_\mathrm{B} = f_\mathrm{B}^{-1} \ll \tau_\mathrm{life}$. It means that magnetic fields oscillations can happen multiple times within the star lifetime. Hence, although a dense axion star is quasi-stable, it provides a reasonable initial condition for the studies of oscillating magnetic fields driven by axions.

We take the magnetic diffusion coefficient as $\eta=10^{10}\,\text{cm}^{2}\cdot\text{s}^{-1}$,
which is close to the observed value and corresponds to the turbulent
magnetic diffusion~\cite{Cha08}. In this case, using the chosen axions energy density in the star, one has that $\phi_{1}(0)=2\times10^{-5}$,
$\phi_{2,\dots,4}(0)=0$, and $\dot{\phi}_{1,\dots,4}(0)=0$.

We suppose that the seed magnetic field is $B(0)=1\,\text{G}$, which
corresponds to the strength in the quiet Sun. We study the situations
of both poloidal and toroidal seed fields. In the former case, $a_{1}(0)=1.4\times10^{-21}$,
$a_{2,\dots,4}(0)=0$, and $b_{1,\dots,4}(0)=0$ for the chosen parameters
of the system. In the latter case, one has that $b_{1}(0)=1.4\times10^{-21}$,
$b_{2,\dots,4}(0)=0$, and $a_{1,\dots,4}(0)=0$. More detailed justification
of the choice of the initial initial condition and the parameters
of an axion is present in Ref.~\cite{Dvo24}.

\section{Results\label{sec:RES}}

\begin{figure}
  \centering
  \subfigure[]
  {\label{fig:f1a}
  \includegraphics[scale=.35]{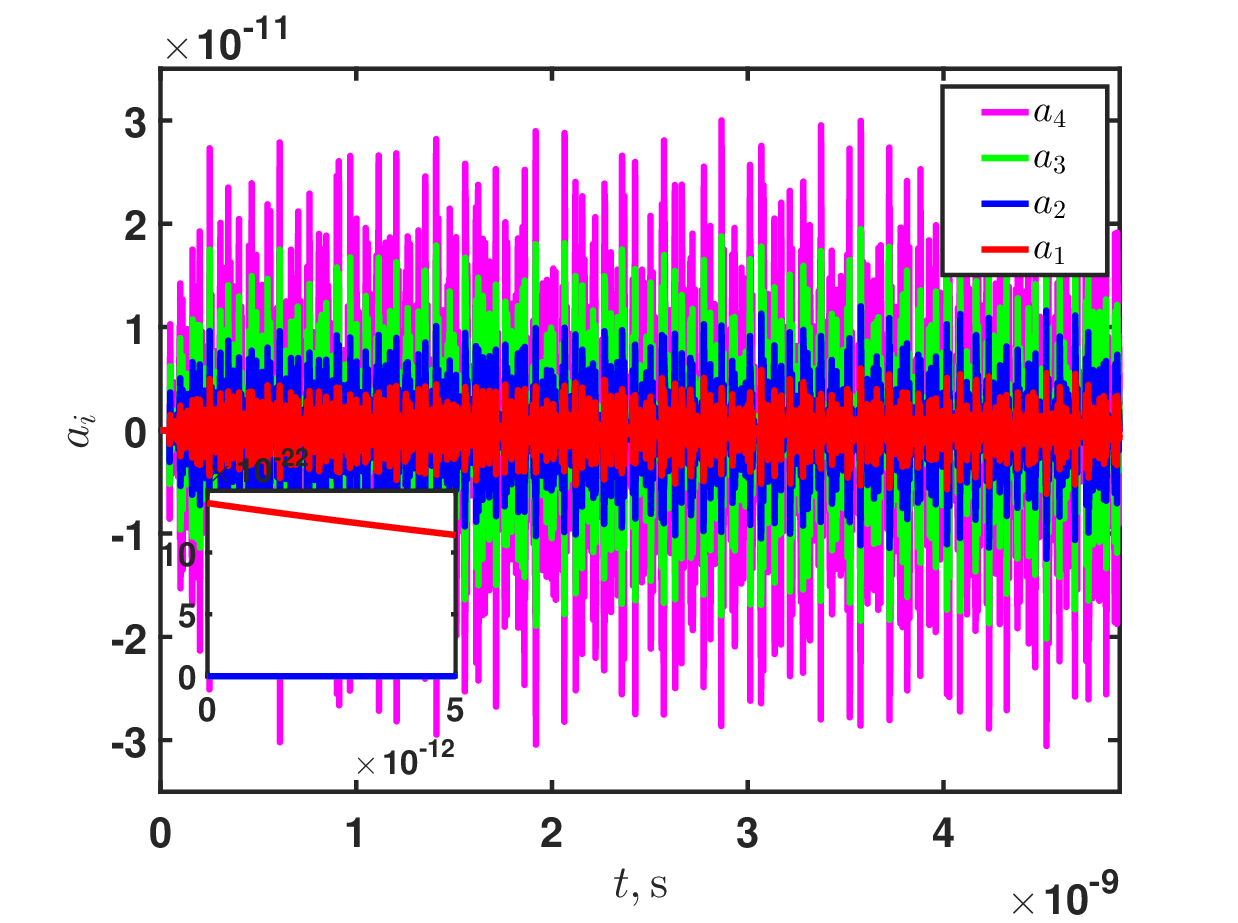}}
  \subfigure[]
  {\label{fig:f1b}
  \includegraphics[scale=.35]{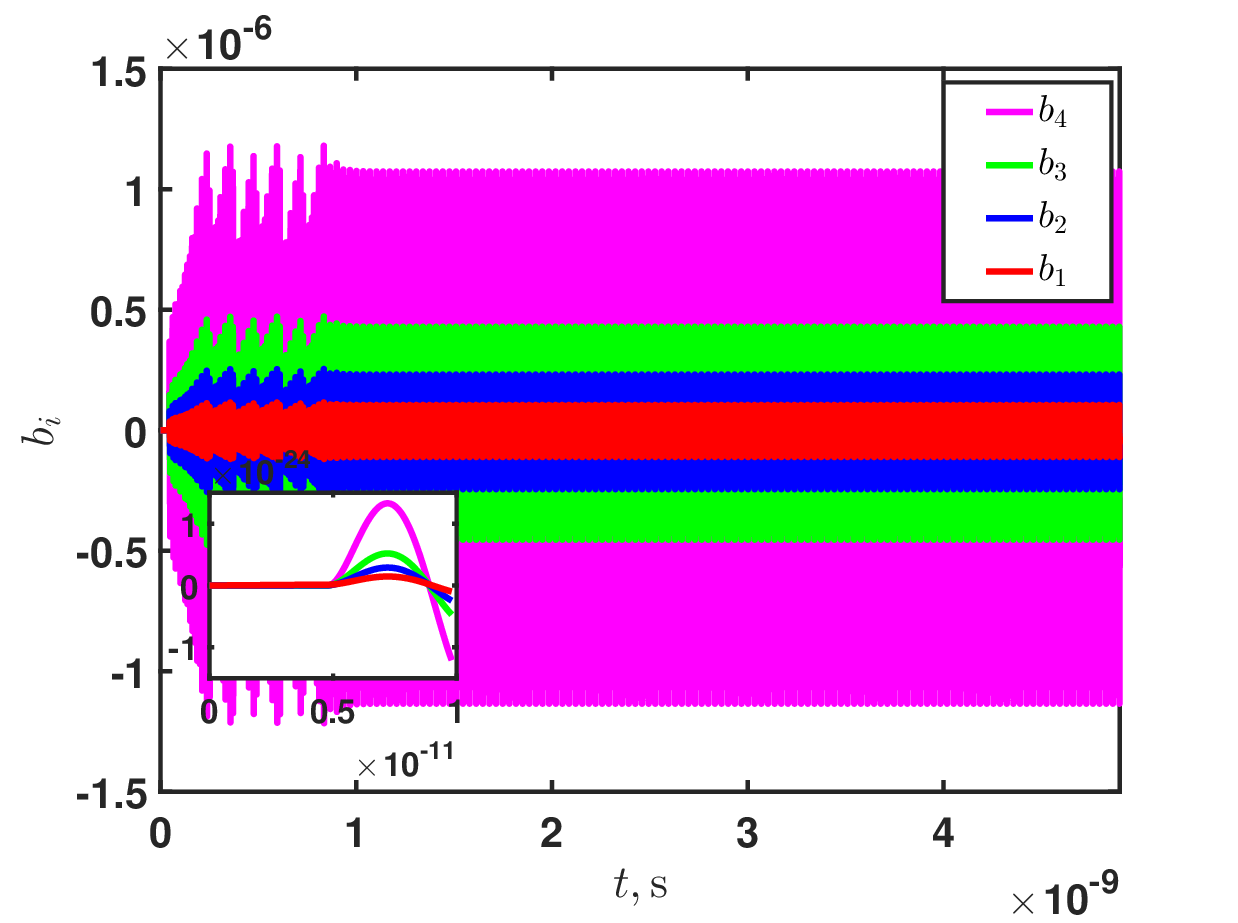}}
  \\
  \subfigure[]
  {\label{fig:f1c}
  \includegraphics[scale=.35]{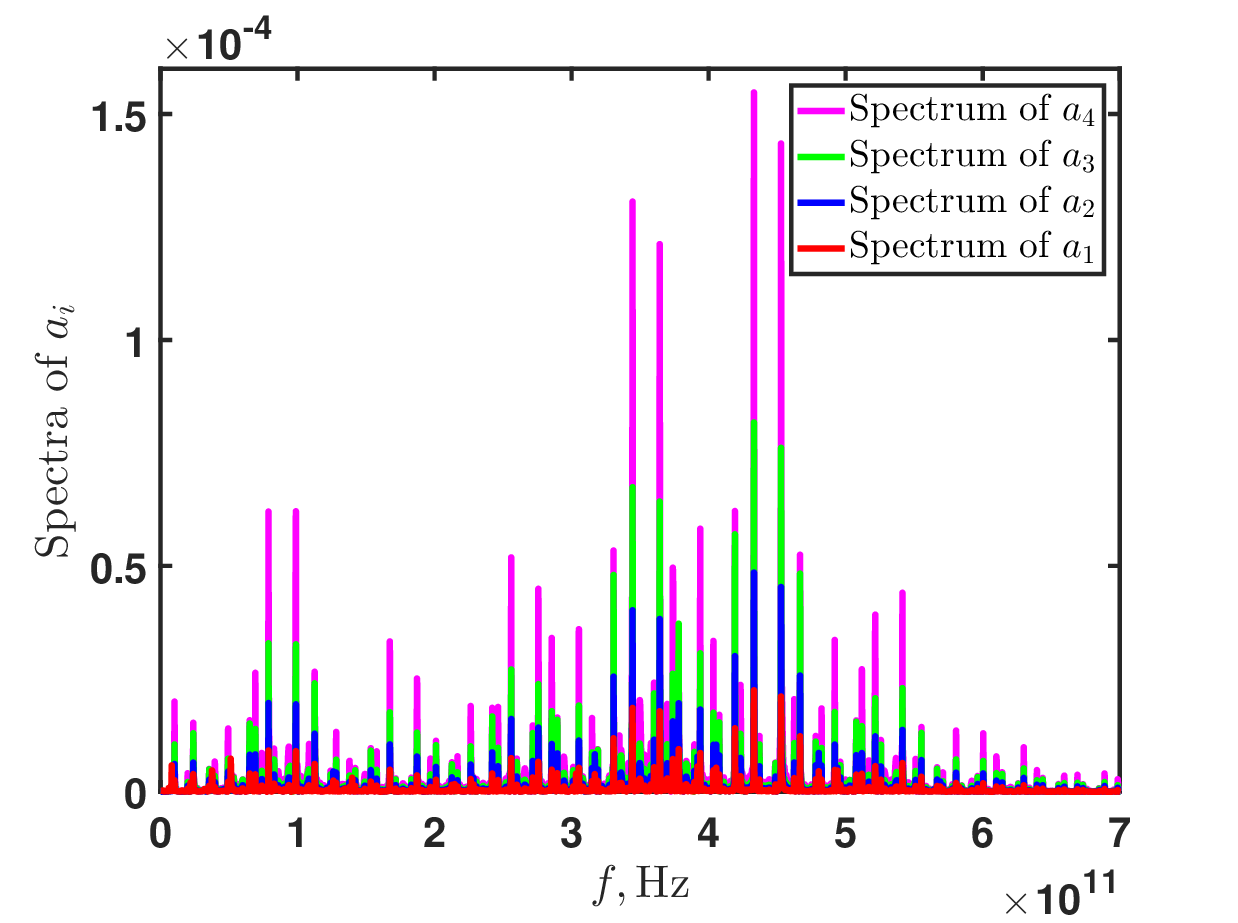}}
  \subfigure[]
  {\label{fig:f1d}
  \includegraphics[scale=.35]{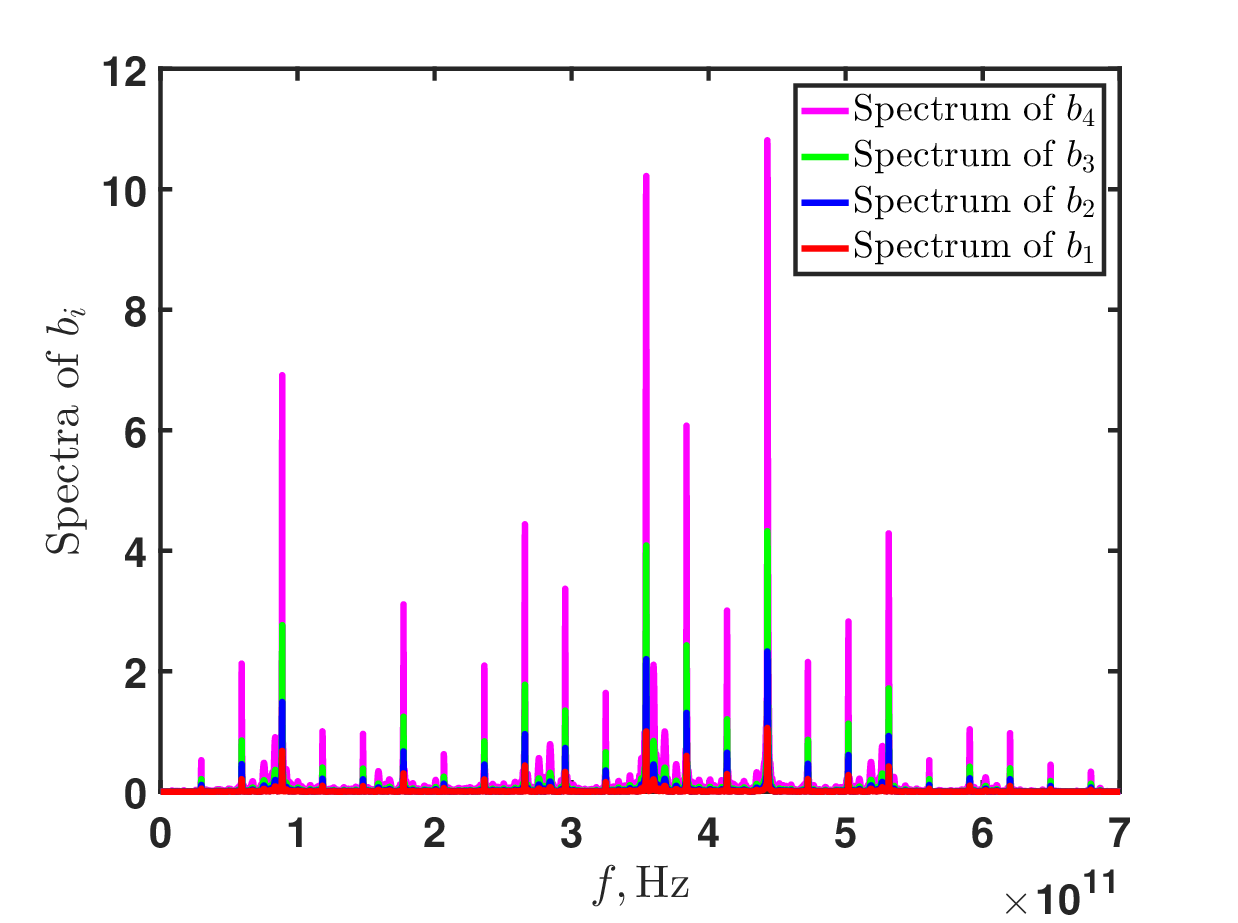}}
  \protect
\caption{Evolution of the magnetic fields harmonics in a dense axion star when
a seed poloidal field is present. The parameters of the system are
the following: the axion mass is $m=10^{-5}\,\text{eV}$, the axion
clump radius is $R=0.7\,\text{cm}$, the seed poloidal magnetic field
is $B_{\mathrm{pol}}(0)=1\,\text{G}$, whereas $B_{\mathrm{tor}}(0)=0$.
(a) The time evolution of the poloidal harmonics. (b) The behavior
of toroidal harmonics. (c) The spectra of poloidal harmonics. (d)
The spectra of toroidal harmonics. The insets in panels~(a) and~(b)
show the evolution of harmonics on short time intervals.\label{fig:poloidal}}
\end{figure}

\begin{figure}
  \centering
  \subfigure[]
  {\label{fig:f2a}
  \includegraphics[scale=.35]{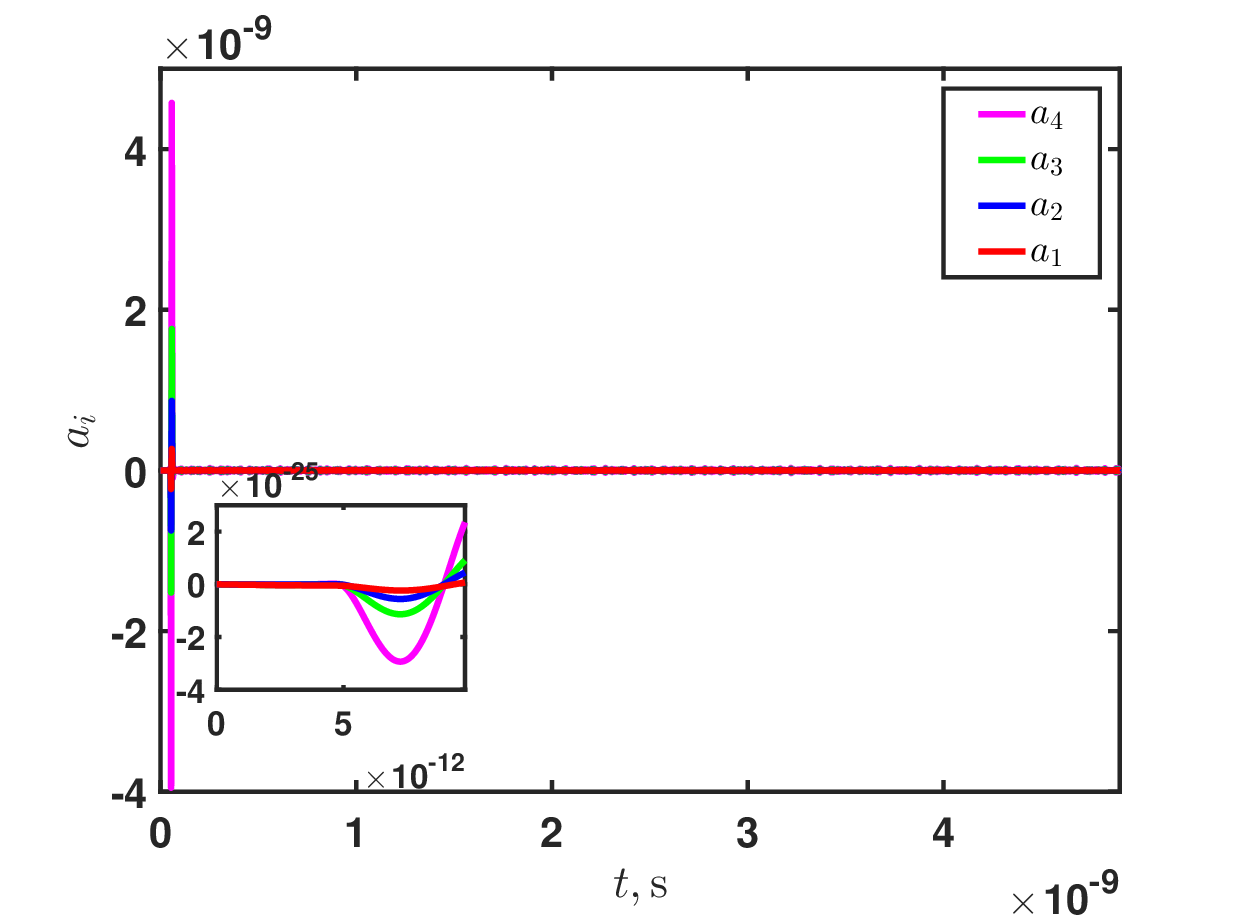}}
  \subfigure[]
  {\label{fig:f2b}
  \includegraphics[scale=.35]{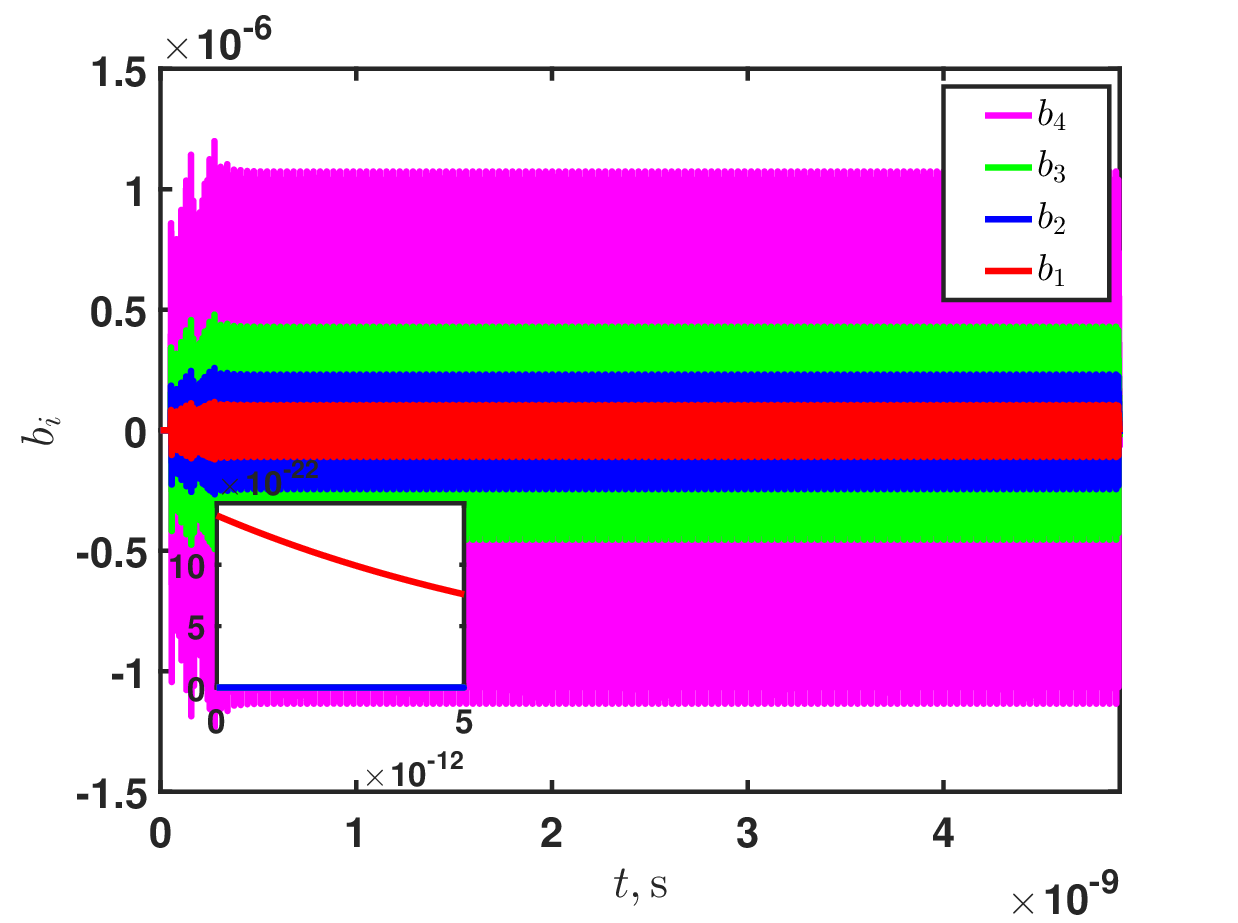}}
  \\
  \subfigure[]
  {\label{fig:f2c}
  \includegraphics[scale=.35]{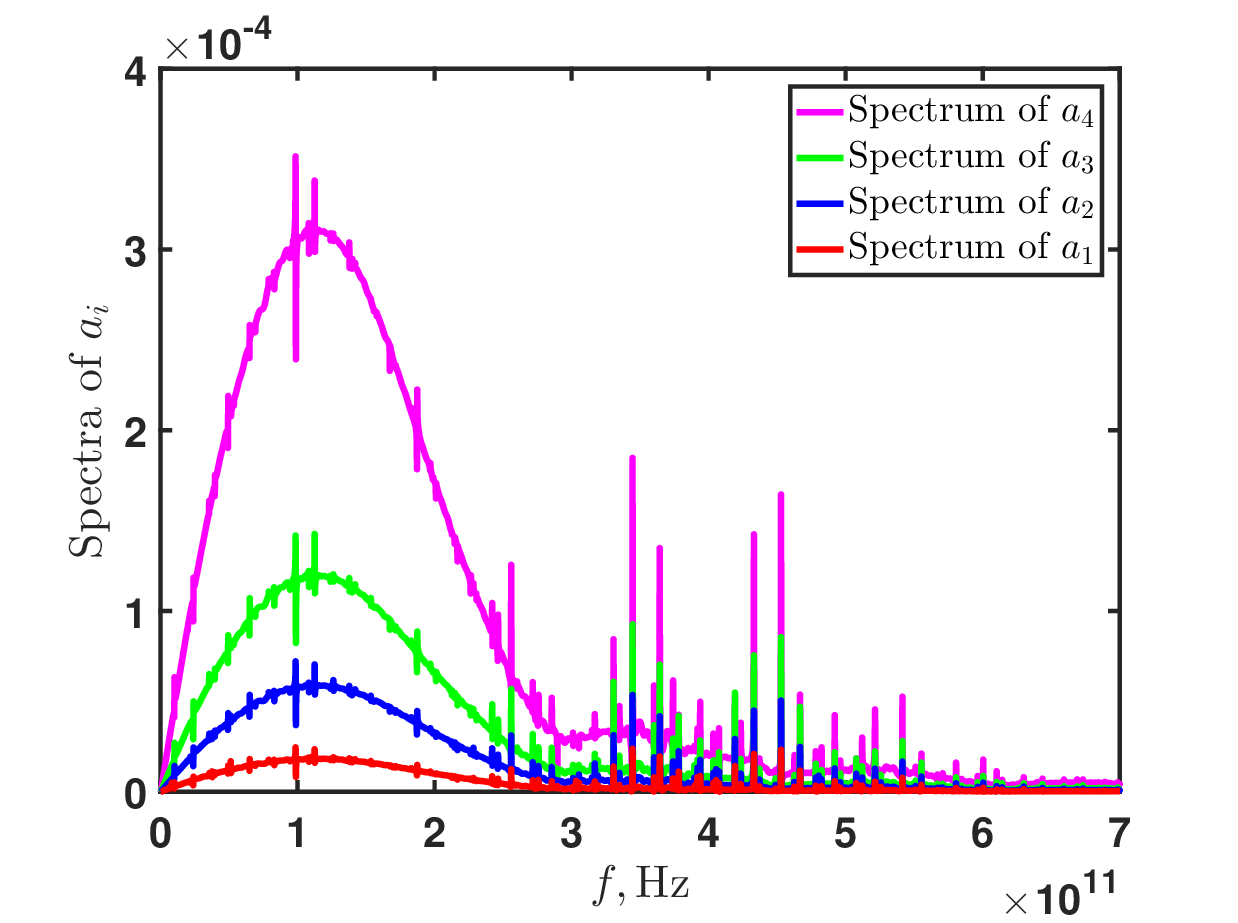}}
  \subfigure[]
  {\label{fig:f2d}
  \includegraphics[scale=.35]{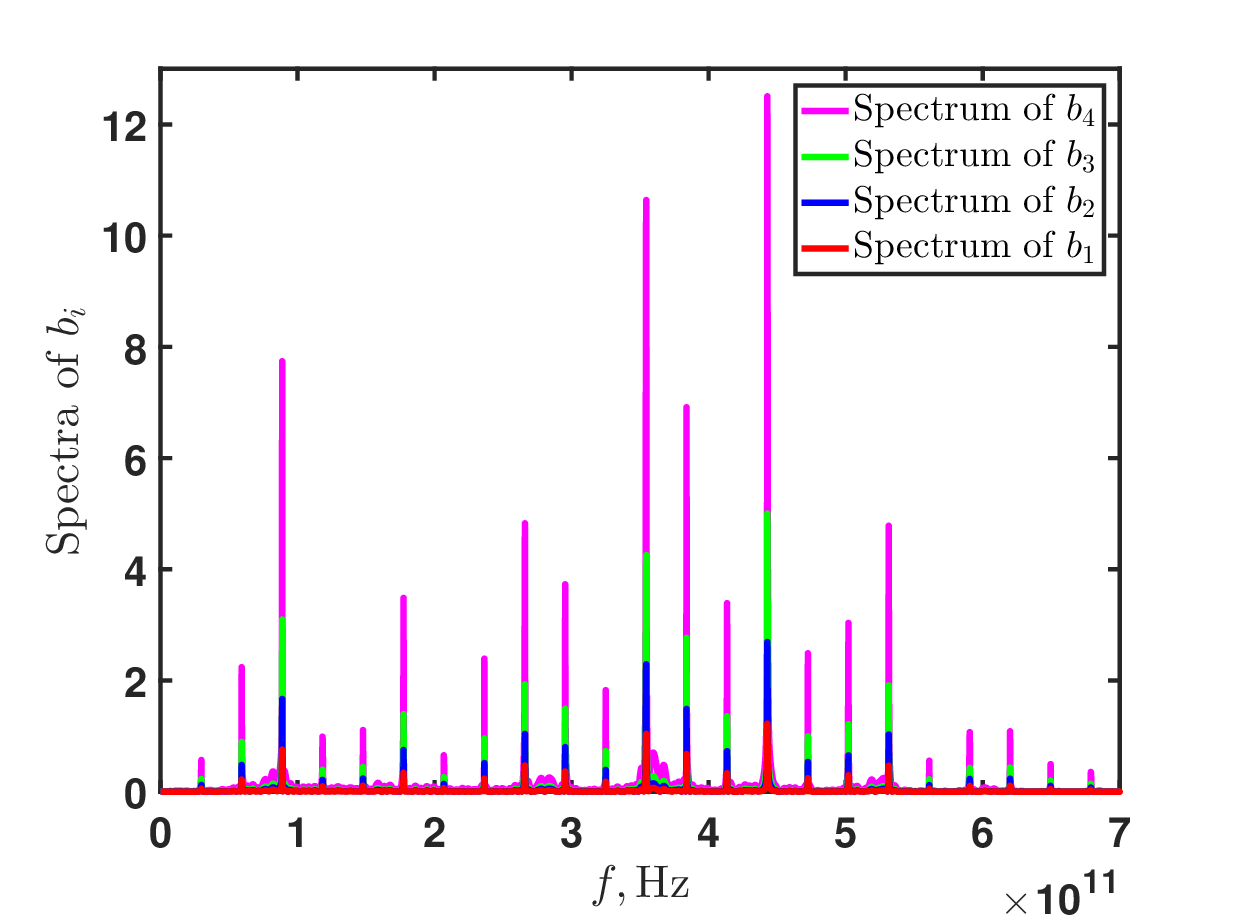}}
  \protect
\caption{The same as in Fig.~\ref{fig:poloidal} for a seed toroidal field,
$B_{\mathrm{tor}}(0)=1\,\text{G}$ and $B_{\mathrm{pol}}(0)=0$.\label{fig:toroidal}}
\end{figure}

In this section, we present the results of the numerical solution
of Eq.~(\ref{eq:abpsieq}) for the initial condition specified in
Sec.~\ref{sec:AXMHD}. In Figs.~\ref{fig:f1a}, \ref{fig:f1b},
\ref{fig:f2a}, and~\ref{fig:f2b}, one can see the
time evolution of the harmonics of poloidal $a_{i}$ and toroidal
$b_{i}$ fields for the cases of either poloidal or toroidal seed
fields. It can be seen in Figs.~\ref{fig:f1b} and~\ref{fig:f2b}
that oscillations of the toroidal harmonics are excited in both cases.
However, the oscillating poloidal field exists only when one has
a seed poloidal field; cf. Figs.~\ref{fig:f1a} and~\ref{fig:f2a}.
It is the new feature compared to Ref.~\cite{Dvo24}. The insets
in Figs.~\ref{fig:f1a}, \ref{fig:f1b}, \ref{fig:f2a},
and~\ref{fig:f2b} demonstrate that the initial condition
is satisfied for $a_{i}$ and $b_{i}$.

Moreover, the next interesting feature to be noticed in Figs.~\ref{fig:f1a},
\ref{fig:f1b}, and~\ref{fig:f2b} is that the amplitudes
of oscillations increases for higher harmonics. It can be the indication
that magnetic field is generically unstable in presence of oscillating
inhomogeneous axions. This feature can be also observed in Figs.~\ref{fig:f1c},
\ref{fig:f1d}, and~\ref{fig:f2d} where the spectra
of harmonics are shown. Note that we do not discuss the spectra of
$a_{i}$ in Fig.~\ref{fig:f2c} which are mostly continuous
since oscillations of poloidal harmonics are not excited. We also
mention in Figs.~\ref{fig:f1c}, \ref{fig:f1d},
and~\ref{fig:f2d} that typical oscillations frequencies
are higher than those in the two harmonics case considered in Ref.~\cite{Dvo24}.

\begin{figure}
  \centering
  \subfigure[]
  {\label{fig:f3a}
  \includegraphics[scale=.35]{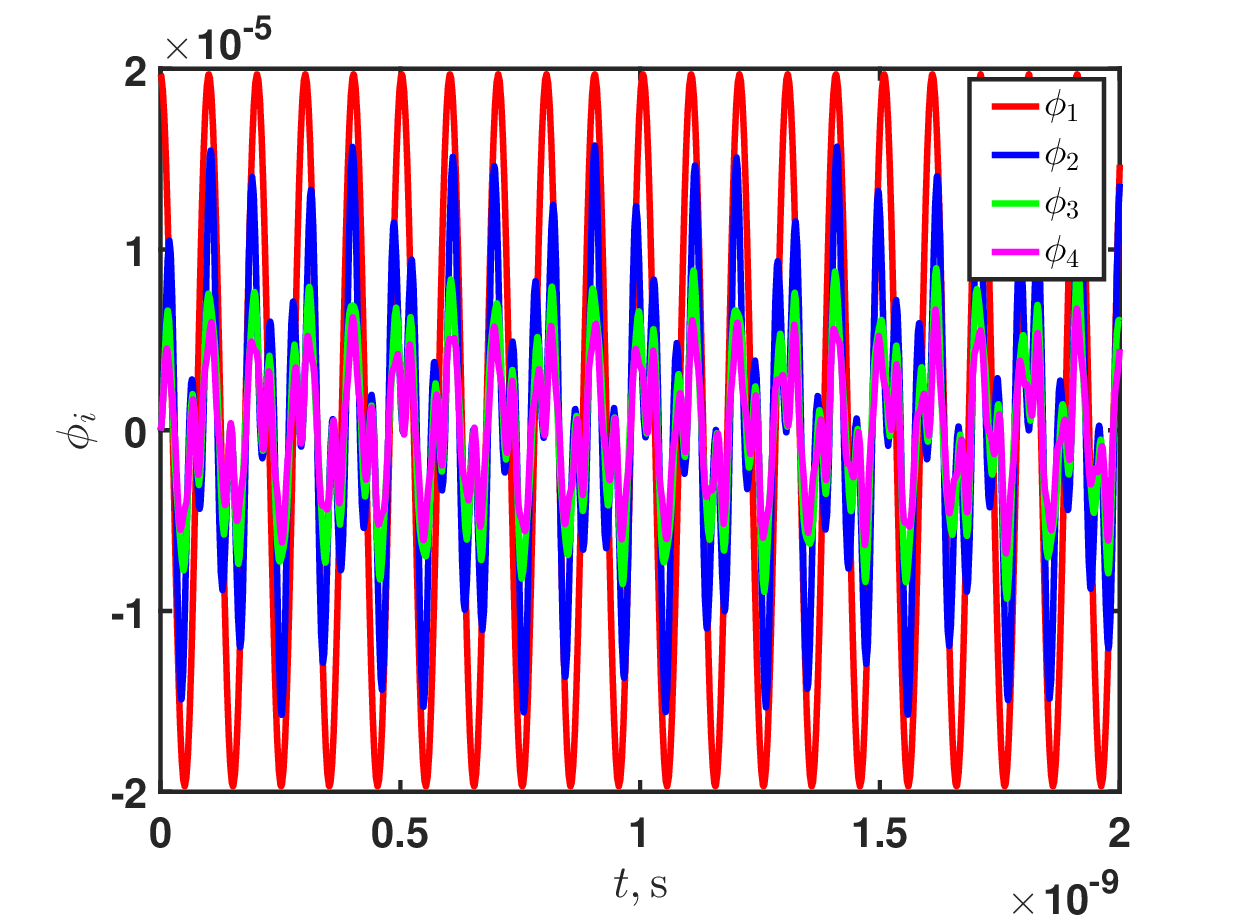}}
  \subfigure[]
  {\label{fig:f3b}
  \includegraphics[scale=.35]{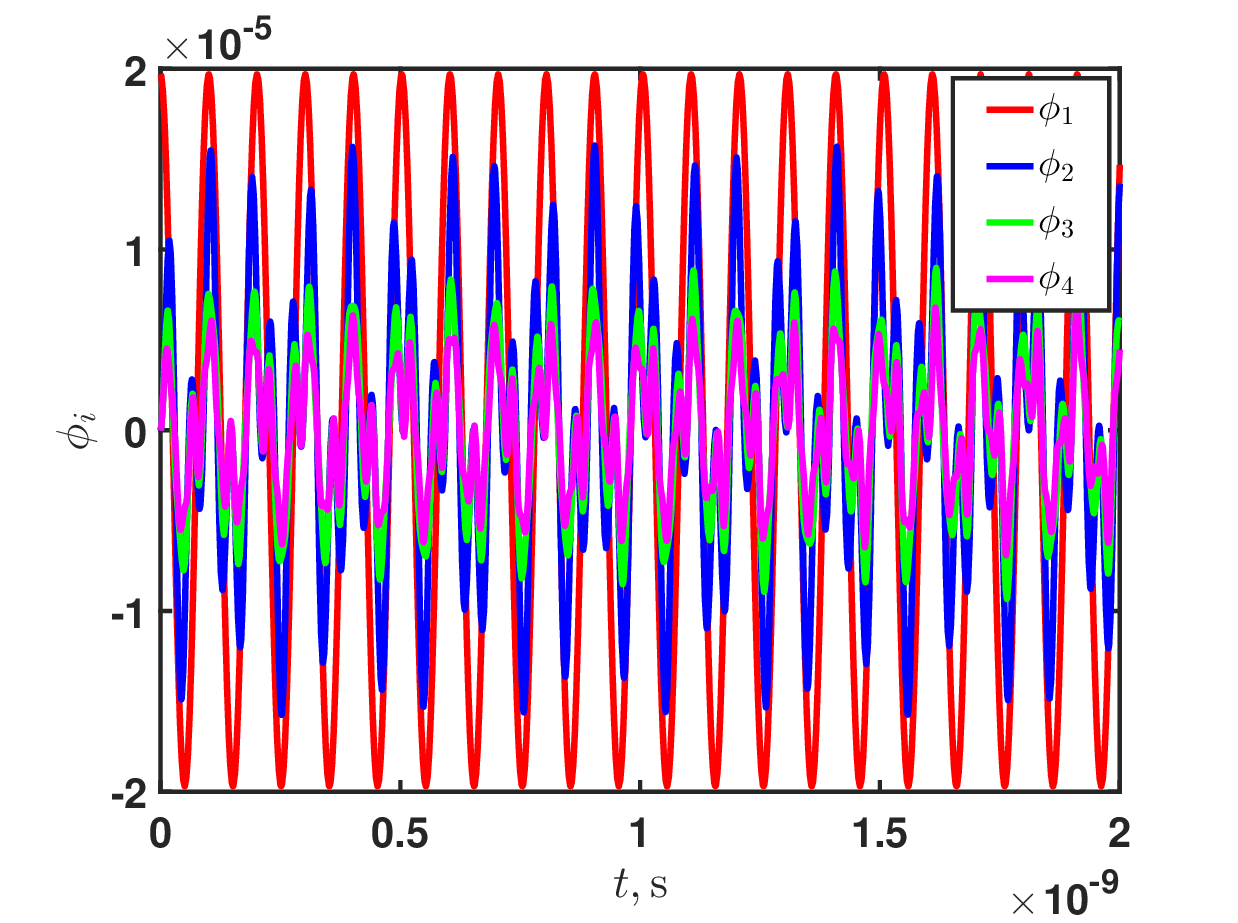}}
  \protect
\caption{The evolution of the axion field harmonics for (a) the seed poloidal
magnetic field and (b) the seed toroidal field. The parameters of
the system and the initial condition correspond to Figs.~\ref{fig:poloidal}
and~\ref{fig:toroidal}.\label{fig:axions}}
\end{figure}

Finally, in Fig.~\ref{fig:axions}, we depict the time evolution
of the axion field harmonics $\phi_{i}$ for both poloidal and toroidal
seed magnetic fields. First, one can see that the evolution of axions
is almost unaffected by the magnetic field; cf. Figs.~\ref{fig:f3a}
and~\ref{fig:f3b}. This feature was mentioned in Refs.~\cite{DvoAkh24,AkhDvo24}.
Second, the amplitudes of oscillating harmonics decrease for the higher
harmonics numbers. It means that oscillations of axions are likely
to be stable. Of course, the stability of axions oscillations is valid while $t \ll \tau_\mathrm{life}$.

\section{Conclusion\label{sec:CONCL}}

In the present work, we have studied the evolution of magnetic fields
driven by time dependent and spatially inhomogeneous axion fields.
Based on the recently derived induction equation which accounts for
the axions contribution, we have considered a magnetized axion clump
embedded in solar plasma. The mutual evolution of axions and magnetic
fields has been examined in a thin layer within the low mode approximation.
It allowed us to derive the system of nonlinear ordinary differential
equations for the amplitudes of the harmonics. This system has been
solved numerically in the case of a dense axion star, having a small
radius, surrounded by solar plasma with realistic characteristics.

Contrary to Ref.~\cite{Dvo24}, where only two harmonics were taken
into account, now we have considered up to four modes in Eq.~(\ref{eq:ABPhiharm}).
The extension of the decomposition over the zonal harmonics is motivated
by the result in Ref.~\cite{Obr23} that higher harmonics can play
a crucial role in dynamo problems.

The consideration of higher harmonics enabled us to reveal several
interesting features in the axion MHD. First, we have noticed that
different seed magnetic fields result in the different subsequent
magnetic fields evolution. We have obtained that oscillations of both
poloidal and toroidal fields are excited only for a seed poloidal
field. Thus, such an initial magnetic fields configuration seems to
be more stable. Indeed, as shown in Ref.~\cite{BraNor06} a long
lived astrophysical magnetic field should have both poloidal and toroidal components.

Second, we have obtained that higher harmonics of the magnetic field
have greater amplitudes. In some situations, this feature can be noticed
in the two modes case considered in Ref.~\cite{Dvo24}. However,
now this fact can be seen with a naked eye; cf. Figs.~\ref{fig:poloidal}
and~\ref{fig:toroidal}. It can be interpreted as the indication that a magnetic field
instability is present in MHD with inhomogeneous axions. Revealing
the source of such an instability is an interesting and important
problem which requires a separate special attention.

We also mention that we have confirmed our previous result in Refs.~\cite{DvoAkh24,AkhDvo24}
that axions almost are not affected by the magnetic fields. Moreover
the axion field harmonics with higher numbers have decreasing amplitudes
contrary to magnetic ones. Thus, axions oscillations are likely to
be quasi-stable.

We have considered a small axionic clump with a relatively high energy density. It corresponds to a dense axion star described in Ref.~\cite{Bra16}. Such an axion star was shown in Ref.~\cite{Vic18} to be quasi-stable. It decays after $\tau_\mathrm{life} \sim 10^{-8}\,\text{s}$~\cite{Vic18} because of the axions emission~\cite{Lev22}. However, the time scale of magnetic fields oscillations, predicted in our work, is much shorter than $\tau_\mathrm{life}$. Thus, we may neglect the instability of a dense axion star while studying the excitation of magnetic fields oscillations.

In our work, we do not discuss the formation of dense magnetized axion stars. Taking into account their short lifetime, they cannot have a cosmological origin. Nonetheless, axions are reported in Ref.~\cite{DilZio03} to be able to concentrate around usual stars, like the Sun. The existence of quite exotic axionic objects, e.g., axion quark nuggets, in the Sun was suggested in Ref.~\cite{Zhi17}. Perhaps, dense magnetized axion stars can also arise in stellar interiors. However, this issue should be studied separately.

The results of the present work can have the implication to the problem
of the solar corona heating as suggested in Ref.~\cite{Dvo24}. Previously,
the involvement of dark matter to the solar corona heating and to
the radio emission of the Sun was discussed in Refs.~\cite{Zhi17,Ge20}.

\section*{Acknowledgments}

I am indebted to D.~D.~Sokoloff for useful comments.

\appendix

\section{Ordinary differential equations for harmonics\label{sec:ODEHARM}}

The basis functions in Eq.~(\ref{eq:ABPhiharm}) obey the orthogonality
condition,
\begin{equation}\label{eq:orthcond}
  \frac{2}{\pi}
  \int_{0}^{\pi}\sin (n\vartheta)\sin (k\vartheta)\mathrm{d}\vartheta=
  \delta_{nk}.
\end{equation}
Using Eq.~(\ref{eq:orthcond}), we can straightforwardly derive the
following system of ordinary differential equations for the functions
$a_{k}$, $b_{k}$, and $\phi_{k}$:
\begin{align}\label{eq:abpsieq}
  \dot{a}_{1}= & -2(a_{1}+a_{2}+a_{3}+a_{4})+\frac{2}{\pi}
  \bigg(-\frac{16}{315}b_{1}\psi_{3}-\frac{64}{3465}b_{1}\psi_{4}
    -\frac{128}{2145}b_{2}\psi_{4} -\frac{16}{315}b_{3}\psi_{1}
    \nonumber
    \displaybreak[2]
    \\
    &-\frac{64}{3465}b_{4}\psi_{1}-\frac{128}{2145}b_{4}\psi_{2}
    -\frac{64}{15}b_{1}\phi_{1}-\frac{1024}{63}b_{2}\phi_{2}
    -\frac{5184}{143}b_{3}\phi_{3} -\frac{16384}{255}b_{4}\phi_{4}
    \nonumber
    \displaybreak[2]
    \\
    &+\frac{16}{15}b_{1}\psi_{1}+\frac{64}{63}b_{2}\psi_{2}
    +\frac{144}{143}b_{3}\psi_{3}+\frac{256}{255}b_{4}\psi_{4}
    -\frac{64}{195}b_{3}\psi_{4}-\frac{64}{195}b_{4}\psi_{3}
    -\frac{32}{99}b_{2}\psi_{3}
    \nonumber
    \displaybreak[2]
    \\
    & -\frac{32}{99}b_{3}\psi_{2}+\frac{4096}{195}b_{3}\phi_{4}
    +\frac{768}{65}b_{4}\phi_{3}
    -\frac{32}{105}b_{1}\psi_{2}-\frac{32}{105}b_{2}\psi_{1}
    +\frac{256}{3465}b_{4}\phi_{1} +\frac{128}{11}b_{2}\phi_{3}
    \nonumber
    \displaybreak[2]
    \\
    &+\frac{512}{99}b_{3}\phi_{2}+\frac{8192}{2145}b_{2}\phi_{4}
    +\frac{2048}{2145}b_{4}\phi_{2}
    +\frac{512}{105}b_{1}\phi_{2}+\frac{128}{105}b_{2}\phi_{1}
    +\frac{64}{35}b_{1}\phi_{3}+\frac{64}{315}b_{3}\phi_{1}
    \nonumber
    \displaybreak[2]
    \\
    & +\frac{4096}{3465}b_{1}\phi_{4}
  \bigg),
  \nonumber
  \displaybreak[2]
  \\
  \dot{a}_{2}= & -6(2a_{2}+a_{3}+a_{4})+\frac{2}{\pi}
  \bigg(
    -\frac{144}{385}b_{1}\psi_{3}-\frac{64}{819}b_{1}\psi_{4}
    -\frac{128}{315}b_{2}\psi_{4}-\frac{144}{385}b_{3}\psi_{1}
    \nonumber
    \displaybreak[2]
    \\
    & -\frac{64}{819}b_{4}\psi_{1}-\frac{128}{315}b_{4}\psi_{2}
    -\frac{64}{35}b_{1}\phi_{1}-\frac{20992}{3465}b_{2}\phi_{2}
    -\frac{27328}{2145}b_{3}\phi_{3}-\frac{1390592}{62985}b_{4}\phi_{4}
    \nonumber
    \displaybreak[2]
    \\
    & +\frac{16}{21}b_{1}\psi_{1} +\frac{64}{165}b_{2}\psi_{2}
    +\frac{16}{45}b_{3}\psi_{3}+\frac{256}{741}b_{4}\psi_{4}
    +\frac{576}{935}b_{3}\psi_{4}+\frac{576}{935}b_{4}\psi_{3}
    +\frac{288}{455}b_{2}\psi_{3}
    \nonumber
    \displaybreak[2]
    \\
    & + \frac{288}{455}b_{3}\psi_{2}-\frac{813568}{36465}b_{3}\phi_{4}
    -\frac{424704}{12155}b_{4}\phi_{3}
    +\frac{32}{45}b_{1}\psi_{2}+\frac{32}{45}b_{2}\psi_{1}
    +\frac{24832}{45045}b_{4}\phi_{1}
    \nonumber
    \displaybreak[2]
    \\
    & - \frac{148352}{15015}b_{2}\phi_{3}
    - \frac{837376}{45045}b_{3}\phi_{2}
    + \frac{318464}{15015} b_{2}\phi_{4}
    + \frac{80896}{9009} b_{4}\phi_{2} - \frac{256}{105}b_{1}\phi_{2}
    \nonumber
    \displaybreak[2]
    \\
    & -\frac{2176}{315}b_{2}\phi_{1}+\frac{11584}{1155}b_{1}\phi_{3}
    +\frac{9664}{3465}b_{3}\phi_{1}+\frac{65024}{15015}b_{1}\phi_{4}
  \bigg),
  \nonumber
  \displaybreak[2]
  \\
  \dot{a}_{3}= & -10(3a_{3}+a_{4})
  +\frac{2}{\pi}
  \bigg(
    \frac{80}{117}b_{1}\psi_{3}
    -\frac{64}{165}b_{1}\psi_{4}+\frac{640}{1071}b_{2}\psi_{4}
    +\frac{80}{117}b_{3}\psi_{1}-\frac{64}{165}b_{4}\psi_{1}
    \nonumber
    \displaybreak[2]
    \\
    & 
    +\frac{640}{1071}b_{4}\psi_{2}
    -\frac{192}{35}b_{1}\phi_{1}-\frac{37376}{9009}b_{2}\phi_{2}
    -\frac{2135488}{255255}b_{3}\phi_{3}
    -\frac{13588480}{969969}b_{4}\phi_{4}-\frac{16}{45}b_{1}\psi_{1}
    \nonumber
    \displaybreak[2]
    \\
    & 
    +\frac{64}{195}b_{2}\psi_{2}
    +\frac{144}{595}b_{3}\psi_{3}+\frac{256}{1155}b_{4}\psi_{4}
    +\frac{320}{1197}b_{3}\psi_{4}
    +\frac{320}{1197}b_{4}\psi_{3}+\frac{32}{105}b_{2}\psi_{3}
    +\frac{32}{105}b_{3}\psi_{2}
    \nonumber
    \displaybreak[2]
    \\
    & -\frac{138538496}{14549535}b_{3}\phi_{4}
    -\frac{72285952}{4849845}b_{4}\phi_{3}
    +\frac{160}{231}b_{1}\psi_{2}+\frac{160}{231}b_{2}\psi_{1}
    +\frac{18688}{4095}b_{4}\phi_{1}-\frac{14464}{3003}b_{2}\phi_{3}
    \nonumber
    \displaybreak[2]
    \\
    & -\frac{364288}{45045}b_{3}\phi_{2}
    -\frac{7773184}{765765}b_{2}\phi_{4}
    -\frac{17724416}{765765}b_{4}\phi_{2}
    -\frac{1280}{693}b_{1}\phi_{2}-\frac{896}{495}b_{2}\phi_{1}
    -\frac{320}{143}b_{1}\phi_{3}
    \nonumber
    \displaybreak[2]
    \\
    & -\frac{45504}{5005}b_{3}\phi_{1}+\frac{31232}{2145}b_{1}\phi_{4} 
  \bigg),
  \nonumber
  \displaybreak[2]
  \\
  \dot{a}_{4}= & -56a_{4}+\frac{2}{\pi}
  \bigg(
    \frac{112}{165}b_{1}\psi_{3}+\frac{448}{663}b_{1}\psi_{4}
    +\frac{896}{3135}b_{2}\psi_{4}+\frac{112}{165}b_{3}\psi_{1}
    +\frac{448}{663}b_{4}\psi_{1}+\frac{896}{3135}b_{4}\psi_{2}
    \nonumber
    \displaybreak[2]
    \\
    & -\frac{9536}{3465}b_{1}\phi_{1}-\frac{34816}{45045}b_{2}\phi_{2}
    -\frac{4525504}{692835}b_{3}\phi_{3}
    -\frac{3595882496}{334639305}b_{4}\phi_{4}
    -\frac{16}{231}b_{1}\psi_{1}
    \nonumber
    \displaybreak[2]
    \\
    & +\frac{64}{105}b_{2}\psi_{2}+\frac{144}{665}b_{3}\psi_{3}
    +\frac{256}{1449}b_{4}\psi_{4}+\frac{64}{315}b_{3}\psi_{4}
    +\frac{64}{315}b_{4}\psi_{3}+\frac{224}{765}b_{2}\psi_{3}
    \nonumber
    \displaybreak[2]
    \\
    & +\frac{224}{765}b_{3}\psi_{2}
    -\frac{14990336}{2078505}b_{3}\phi_{4}
    -\frac{49881856}{4849845}b_{4}\phi_{3}-\frac{224}{585}b_{1}\psi_{2}
    -\frac{224}{585}b_{2}\psi_{1}-\frac{8440064}{765765}b_{4}\phi_{1}
    \nonumber 
    \displaybreak[2]
    \\
    & -\frac{144256}{36465}b_{2}\phi_{3}
    -\frac{1319936}{255255}b_{3}\phi_{2}
    -\frac{3110912}{692835}b_{2}\phi_{4}
    -\frac{12947456}{1322685}b_{4}\phi_{2}
    -\frac{7168}{715}b_{1}\phi_{2}
    \nonumber
    \displaybreak[2]
    \\
    & -\frac{404608}{45045}b_{2}\phi_{1}-\frac{1344}{715}b_{1}\phi_{3}
    -\frac{56512}{45045}b_{3}\phi_{1}-\frac{7168}{3315}b_{1}\phi_{4}
  \bigg),
  \nonumber
  \displaybreak[2]
  \\
  \dot{b}_{1}= & -2(3b_{1}+2b_{2}+2b_{3}+2b_{4})+\frac{2}{\pi}
  \bigg(
    \frac{1757968}{15015}a_{3}\phi_{3}
    -\frac{1301824}{45045}a_{2}\phi_{4}
    -\frac{11948224}{45045}a_{3}\phi_{4}
    \nonumber
    \displaybreak[2]
    \\
    & -\frac{164576}{45045}a_{4}\psi_{2}+\frac{304}{105}a_{2}\phi_{1}
    +\frac{10064}{105}a_{3}\phi_{1}+\frac{16}{15}a_{1}\psi_{1}
    +\frac{112}{15}a_{1}\phi_{1}+\frac{90800}{9009}a_{4}\psi_{3}
    \nonumber
    \displaybreak[2]
    \\
    & +\frac{678464}{765765}a_{4}\psi_{4}+\frac{608}{315}a_{2}\psi_{2}
    +\frac{27872}{3465}a_{3}\psi_{2}
    +\frac{25328}{3465}a_{2}\psi_{3}+\frac{70736}{45045}a_{3}\psi_{3}
    \nonumber
    \displaybreak[2]
    \\
    & +\frac{18752}{5005}a_{2}\psi_{4}+\frac{95168}{9009}a_{3}\psi_{4}
    +\frac{2804320}{9009}a_{4}\phi_{2}
    -\frac{783472}{5005}a_{4}\phi_{3}
    +\frac{199878592}{765765}a_{4}\phi_{4}
    \nonumber
    \displaybreak[2]
    \\
    & +\frac{288}{7}a_{2}\phi_{2}-\frac{39584}{1155}a_{3}\phi_{2}
    -\frac{16528}{231}a_{2}\phi_{3}+\frac{352}{105}a_{1}\psi_{2}
    +\frac{496}{315}a_{1}\psi_{3}+\frac{3776}{3465}a_{1}\psi_{4}
    \nonumber
    \displaybreak[2]
    \\
    & -\frac{208}{105}a_{3}\psi_{1}-\frac{1264}{3465}a_{4}\psi_{1}
    +\frac{592}{105}a_{2}\psi_{1}-\frac{608}{105}a_{1}\phi_{2}
    -\frac{208}{105}a_{1}\phi_{3}-\frac{4288}{3465}a_{1}\phi_{4}
    \nonumber
    \displaybreak[2]
    \\
    &+\frac{398192}{3465}a_{4}\phi_{1}
  \bigg),
  \nonumber
  \displaybreak[2]
  \\
  \dot{b}_{2}= & -4(5b_{2}+2b_{3}+2b_{4})
  +\frac{2}{\pi}
  \bigg(
    \frac{2831776}{15015}a_{3}\phi_{3}-\frac{889984}{5005}a_{2}\phi_{4}
    +\frac{34181248}{85085}a_{3}\phi_{4}
    \nonumber
    \displaybreak[2]
    \\
    & +\frac{940864}{45045}a_{4}\psi_{2}+\frac{288}{7}a_{2}\phi_{1}
    -\frac{39584}{1155}a_{3}\phi_{1}
    -\frac{416}{105}a_{1}\psi_{1}-\frac{608}{105}a_{1}\phi_{1}
    +\frac{1374944}{153153}a_{4}\psi_{3}
    \nonumber
    \displaybreak[2]
    \\
    & +\frac{80046976}{14549535}a_{4}\psi_{4}
    +\frac{11584}{3465}a_{2}\psi_{2}+\frac{319936}{45045}a_{3}\psi_{2}
    +\frac{67552}{45045}a_{2}\psi_{3}+\frac{42656}{9009}a_{3}\psi_{3}
    \nonumber
    \displaybreak[2]
    \\
    & +\frac{120448}{9009}a_{2}\psi_{4}
    -\frac{121984}{765765}a_{3}\psi_{4}
    -\frac{9927488}{45045}a_{4}\phi_{2}
    +\frac{11138208}{85085}a_{4}\phi_{3}
    +\frac{4672509056}{14549535}a_{4}\phi_{4}
    \nonumber
    \displaybreak[2]
    \\
    &+\frac{294592}{3465}a_{2}\phi_{2}
    +\frac{4176448}{45045}a_{3}\phi_{2}
    +\frac{2204512}{15015}a_{2}\phi_{3}+\frac{64}{63}a_{1}\psi_{2}
    +\frac{608}{99}a_{1}\psi_{3} +\frac{6016}{2145}a_{1}\psi_{4}
    \nonumber
    \displaybreak[2]
    \\
    &+\frac{11360}{693}a_{3}\psi_{1}-\frac{5152}{715}a_{4}\psi_{1}
    +\frac{1888}{315}a_{2}\psi_{1}+\frac{1984}{63}a_{1}\phi_{2}
   -\frac{544}{33}a_{1}\phi_{3}-\frac{10112}{2145}a_{1}\phi_{4}
    \nonumber
    \displaybreak[2]
    \\
    &+\frac{2804320}{9009}a_{4}\phi_{1}
  \bigg),
  \nonumber
  \displaybreak[2]
  \\
  \dot{b}_{3}= & -6(7b_{3}+2b_{4})+\frac{2}{\pi}
  \bigg(
    \frac{26318352}{85085}a_{3}\phi_{3}+\frac{3693504}{12155}a_{2}\phi_{4}+\frac{141110848}{323323}a_{3}\phi_{4}
    \nonumber
    \displaybreak[2]
  \\
  & +\frac{165536}{12155}a_{4}\psi_{2}-\frac{16528}{231}a_{2}\phi_{1}+\frac{1757968}{15015}a_{3}\phi_{1}-\frac{176}{105}a_{1}\psi_{1}
  -\frac{208}{105}a_{1}\phi_{1}  
    \nonumber
    \displaybreak[2]
  \\
  &+\frac{45331024}{4849845}a_{4}\psi_{3}+\frac{35101376}{4849845}a_{4}\psi_{4}+\frac{81824}{15015}a_{2}\psi_{2}
  +\frac{113504}{15015}a_{3}\psi_{2}
  +\frac{6736}{2145}a_{2}\psi_{3} 
    \nonumber
    \displaybreak[2]
  \\
  &+\frac{1455952}{255255}a_{3}\psi_{3}+\frac{12864}{12155}a_{2}\psi_{4}+\frac{7240384}{1616615}a_{3}\psi_{4}+\frac{11138208}{85085}a_{4}\phi_{2} 
    \nonumber
    \displaybreak[2]
    \\
   & +\frac{541603344}{1616615}a_{4}\phi_{3}
  +\frac{2551364416}{4849845}a_{4}\phi_{4}+\frac{2204512}{15015}a_{2}\phi_{2}+\frac{2831776}{15015}a_{3}\phi_{2}
    \nonumber
    \displaybreak[2]
    \\
    & +\frac{147216}{715}a_{2}\phi_{3} -\frac{224}{33}a_{1}\psi_{2}
    +\frac{144}{143}a_{1}\psi_{3}+\frac{576}{65}a_{1}\psi_{4}+\frac{7120}{429}a_{3}\psi_{1}+\frac{23184}{715}a_{4}\psi_{1}
    \nonumber
    \displaybreak[2]
    \\
    &-\frac{13744}{1155}a_{2}\psi_{1}
    -\frac{544}{33}a_{1}\phi_{2}+\frac{10224}{143}a_{1}\phi_{3}
 -\frac{2112}{65}a_{1}\phi_{4}-\frac{783472}{5005}a_{4}\phi_{1}
  \bigg),
\nonumber
\displaybreak[2]
\\
  \dot{b}_{4}= & -72b_{4}+\frac{2}{\pi}
  \bigg(
    \frac{141110848}{323323}a_{3}\phi_{3}+\frac{23567104}{62985}a_{2}\phi_{4}+\frac{2877294848}{4849845}a_{3}\phi_{4}
    \nonumber
    \displaybreak[2]
    \\
    & +\frac{1929088}{138567}a_{4}\psi_{2}-\frac{1301824}{45045}a_{2}\phi_{1}-\frac{11948224}{45045}a_{3}\phi_{1}
    -\frac{3904}{3465}a_{1}\psi_{1}-\frac{4288}{3465}a_{1}\phi_{1}
    \nonumber
    \displaybreak[2]
    \\
    & +\frac{9557440}{969969}a_{4}\psi_{3}+\frac{2709053696}{334639305}a_{4}\psi_{4}-\frac{827264}{45045}a_{2}\psi_{2}
    +\frac{11891584}{765765}a_{3}\psi_{2}
    \nonumber
    \displaybreak[2]
    \\
    & +\frac{41408}{7293}a_{2}\psi_{3}+\frac{31917632}{4849845}a_{3}\psi_{3}+\frac{193792}{62985}a_{2}\psi_{4}
    +\frac{26019584}{4849845}a_{3}\psi_{4}
    \nonumber
    \displaybreak[2]
    \\
    & +\frac{4672509056}{14549535}a_{4}\phi_{2}
    +\frac{2551364416}{4849845}a_{4}\phi_{3}+\frac{252914761472}{334639305}a_{4}\phi_{4}-\frac{889984}{5005}a_{2}\phi_{2}  
    \nonumber
    \displaybreak[2]
    \\
    & +\frac{34181248}{85085}a_{3}\phi_{2}
    +\frac{3693504}{12155}a_{2}\phi_{3}
    -\frac{6272}{2145}a_{1}\psi_{2}-\frac{1856}{195}a_{1}\psi_{3}+\frac{256}{255}a_{1}\psi_{4}-\frac{48704}{2145}a_{3}\psi_{1}
    \nonumber
    \displaybreak[2]
    \\
    & +\frac{107968}{3315}a_{4}\psi_{1}-\frac{212672}{45045}a_{2}\psi_{1}
    -\frac{10112}{2145}a_{1}\phi_{2}-\frac{2112}{65}a_{1}\phi_{3}+\frac{32512}{255}a_{1}\phi_{4}
    \nonumber
    \displaybreak[2]
    \\
    & +\frac{199878592}{765765}a_{4}\phi_{1}
  \bigg),
  \nonumber
  \displaybreak[2]
  \\
  \dot{\psi}_{1}= & -(\mu^{2}+2\kappa^{2})\phi_{1}
  +\frac{2}{\pi}
  \bigg(
    \frac{112}{15}a_{1}b_{1}-\frac{32}{15}a_{1}b_{2}+\frac{2096}{105}a_{2}b_{1}+\frac{150944}{3465}a_{3}b_{2}
    \nonumber
    \displaybreak[2]
    \\
    & +\frac{1997872}{45045}a_{3}b_{3}-\frac{1189312}{45045}a_{3}b_{4}-\frac{48}{35}a_{4}b_{1}
    -\frac{929696}{45045}a_{4}b_{2}+\frac{3469328}{45045}a_{4}b_{3}
    \nonumber
    \displaybreak[2]
    \\
    & +\frac{59413696}{765765}a_{4}b_{4}-\frac{16}{45}a_{1}b_{3}-\frac{64}{495}a_{1}b_{4}+\frac{6688}{315}a_{2}b_{2}
    -\frac{37232}{3465}a_{2}b_{3}-\frac{33472}{15015}a_{2}b_{4}
    \nonumber
    \displaybreak[2]
    \\
    &-\frac{464}{63}a_{3}b_{1}
  \bigg),
  \nonumber
  \displaybreak[2]
  \\
  \dot{\psi}_{2}= & -(\mu^{2}+12\kappa^{2})\phi_{2}+4\kappa^{2}\phi_{1}
  +\frac{2}{\pi}
  \bigg(
    \frac{544}{105}a_{1}b_{1}+\frac{2752}{315}a_{1}b_{2}+\frac{5408}{315}a_{2}b_{1}+\frac{29504}{1001}a_{3}b_{2}
    \nonumber
    \displaybreak[2]
    \\
    & +\frac{465952}{15015}a_{3}b_{3}
    +\frac{34897024}{765765}a_{3}b_{4}-\frac{769696}{45045}a_{4}b_{1}+\frac{2904256}{45045}a_{4}b_{2}
    +\frac{11538272}{255255}a_{4}b_{3}
    \nonumber
    \displaybreak[2]
    \\
    & + \frac{61287808}{1322685}a_{4}b_{4}
    - \frac{416}{105}a_{1}b_{3}-\frac{36224}{45045}a_{1}b_{4}+\frac{65216}{3465}a_{2}b_{2}
    +\frac{1044128}{45045}a_{2}b_{3}
    \nonumber
    \displaybreak[2]
    \\
    & -\frac{675968}{45045}a_{2}b_{4} + \frac{40672}{1155}a_{3}b_{1}
  \bigg),
  \nonumber
  \displaybreak[2]
  \\
  \dot{\psi}_{3}= & -(\mu^{2}+30\kappa^{2})\phi_{3}+4\kappa^{2}(\phi_{1}+2\phi_{2})
  +\frac{2}{\pi}
  \bigg(
    \frac{304}{105}a_{1}b_{1}+\frac{31072}{3465}a_{1}b_{2} +\frac{1270624}{45045}a_{3}b_{2}
    \nonumber
    \displaybreak[2]
    \\
    & + \frac{9776}{1155}a_{2}b_{1}+\frac{4650064}{153153}a_{3}b_{3}+\frac{95125312}{2909907}a_{3}b_{4}
    +\frac{820912}{15015}a_{4}b_{1} +\frac{1827104}{45045}a_{4}b_{2}
    \nonumber
    \displaybreak[2]
    \\
    & + \frac{53905328}{1322685}a_{4}b_{3}+\frac{617397952}{14549535}a_{4}b_{4}+\frac{429136}{45045}a_{1}b_{3}
    -\frac{271424}{45045}a_{1}b_{4}+\frac{41248}{2145}a_{2}b_{2}
    \nonumber
    \displaybreak[2]
    \\
    & + \frac{312656}{15015}a_{2}b_{3}+\frac{6460352}{255255}a_{2}b_{4}+\frac{33872}{1155}a_{3}b_{1}
  \bigg),
  \nonumber
  \displaybreak[2]
  \\ 
  \dot{\psi}_{4}= & -(\mu^{2}+56\kappa^{2})\phi_{4}+4\kappa^{2}(\phi_{1}+2\phi_{2}+3\phi_{3})
  + \frac{2}{\pi}
  \bigg(
    \frac{7232}{3465}a_{1}b_{1}+\frac{221824}{45045}a_{1}b_{2}  
    \nonumber
    \displaybreak[2]
    \\
    & +\frac{281024}{45045}a_{2}b_{1}
    +\frac{22883456}{765765}a_{3}b_{2}+\frac{8091584}{264537}a_{3}b_{3}+\frac{94617344}{2909907}a_{3}b_{4} 
    +\frac{3905728}{85085}a_{4}b_{1}
    \nonumber
    \displaybreak[2]
    \\
    & +\frac{16716928}{440895}a_{4}b_{2}
    +\frac{115868096}{2909907}a_{4}b_{3}+\frac{14044114688}{334639305}a_{4}b_{4}+\frac{111296}{9009}a_{1}b_{3}
    \nonumber
    \displaybreak[2]
    \\
    & +\frac{7730432}{765765}a_{1}b_{4} +\frac{107648}{6435}a_{2}b_{2}
    + \frac{5281216}{255255}a_{2}b_{3} +\frac{2787584}{124355}a_{2}b_{4}+\frac{34368}{5005}a_{3}b_{1}
  \bigg),
\end{align}
where $\psi_{i}=\dot{\phi}_{i}$ and dot means the derivative with
respect to $\tau$. Note that Eq.~(\ref{eq:abpsieq}) generalizes
analogous system in Ref.~\cite{Dvo24} which was derived when only
two harmonics are accounted for in Eq.~(\ref{eq:ABPhiharm}).

\end{document}